\documentclass[%
 reprint,
 amsmath,amssymb,
 aps,
]{revtex4-2}

\usepackage{graphicx}
\usepackage{dcolumn}
\usepackage{bm}
\usepackage{color}

\begin{document}

\preprint{APS/123-QED}

\title{Dissecting physics of carbon ordering in bcc iron}

\author{Sam Waseda}
 \email{waseda@mpie.de}
\author{Tilmann Hickel}%
\author{Jörg Neugebauer}%
\affiliation{%
 Max-Planck-Institut für Eisenforschung, Max-Planck-Straße 1, D-40237 Düsseldorf, Germany
}%
\author{Julien Morthomas}
\author{Patrice Chantrenne}
\author{Michel Perez}

\affiliation{Univ Lyon, INSA Lyon, UCBL, MATEIS, UMR5510, 69621 Villeurbanne, France}

\date{\today}

\begin{abstract}
Zener ordering is a phenomenon that octahedral interstitial atoms such as carbon occupy the same sublattice inside bcc matrix such as iron. The original formulation relies on a mean field theory, which is still most in use today. We employ multiple methods, such as Molecular Dynamics, Metropolis Monte Carlo, Mean Field Theory with chemical interactions and finite temperature effects to show that the Zener ordering for iron carbon systems is governed by local chemical interactions and finite temperature effects and less of mean field nature as described originally by Zener.
\end{abstract}

\maketitle


\section{Introduction}

Iron-carbon alloys, commonly known as steels, serve as the backbone of modern industrial infrastructure. Understanding the effects of different carbon concentrations and alloying elements on the mechanical properties of steel is essential for producing materials that are strong, ductile, and corrosion-resistant. This knowledge enables engineers to design structures that withstand immense pressures, varying temperatures, and environmental challenges, ensuring safety and longevity in critical applications like bridges, skyscrapers, and pipelines.

In spite of the extensive utilization of steel, which has spanned both a substantial temporal continuum and a significant volumetric scale within human history, full understanding of the energetics as well as the kinetics of carbon inside steels has yet not been attained \cite{zhang2020mechanism}. This deficiency of comprehension arises from the intricate interplay of chemistry, elasticity, and magnetism within the context of iron-carbon as well as carbon-carbon interactions in steels. In ferrite, the anisotropic nature of the three types of sublattice makes certain sublattice alignments of carbon atoms elastically favourable, while it is unfavourable in terms of entropy. Such an alignment is today commonly known as Zener ordering. The understanding of the thermodynamics of Zener ordering will give us insight into the stability of martensite,  the kinetics of carbon atoms at the austenite-ferrite transition, as well as the interplay with defect segregation \cite{zhang2020mechanism}.
 
In its original form, the theory of Zener \cite{zener1948theory}  described the ordering phenomenon on the basis of a mean-field theory (MFT), counting the number of carbon atoms in each of the three sublattices without considering direct interactions between the carbon atoms as well as the temperature effect beyond configurational entropy. Khachaturyan, in his Microscopic Elasticity Theory (MET) \cite{khachaturian1955k,kurdjumov1972phenomena,udyansky2009interplay} included pairwise carbon-carbon interactions, which are repulsive in nature. Waseda \emph{et al.} \cite{waseda2018ordering} compared the ordering phenomenon described by MFT, Metropolis Monte Carlo (MMC) and Molecular Dynamics (MD) and found that short-range carbon-carbon interactions hinder Zener ordering. The combination of the two opposing effects led to the unexpected congruence between the results of MFT and MD at higher temperature \cite{sinclair2010ordering, maugis2018stress,maugis2017temperature, chirkov2015molecular}.
In addition, their results \cite{waseda2018ordering} showed pronounced importance of precipitate-like phases, such as Fe$_{16}$C$_2$. This precipitate structure is commonly observed in systems like iron-nitrogen, but has not been detected for iron-carbon experimentally. The result is also supported by other computational studies that emphasized the importance of short range carbon-carbon interactions leading to the formation of carbides \cite{naraghi2014thermodynamics}. 
These results, however, failed to show the full analysis of finite temperature effects beyond configurational entropy, which was speculated to be the reason why in other works a good agreement between MD and MFT was observed \cite{maugis2017temperature,sinclair2010ordering}. Moreover, according to the MFT it is expected that there is an energy barrier between the ordered and disorderes states, meaning the free energy minimization scheme of MMC might converge to the local minimum and not the true free energy minimum.

In this study, we utilized the metadynamics method \cite{theodoropoulos2000coarse} to understand both long- and short-range interactions of carbon atoms for temperatures ranging from very high values employed in previous studies \cite{maugis2017temperature,sinclair2010ordering}, to below room temperature, which cannot be handled by the conventional MD, as well as varying the carbon content, in order to understand the order-disorder transition. The interpretation of the results is substantiated by the combination of these other methods: MFT alone, MFT with short range carbon-carbon interactions (MFT+CC), MFT+CC with temperature dependent thermodynamic parameters (MFT+CC+FT), conjugated with previous results \cite{waseda2018ordering} employing MMC and classical MD. In order to develop the logic, we first start with the simplest method, namely MFT, and add physical complexity to progressively converge towards the outcomes encompassing the complete intricacies afforded by metadynamics.

\section{Theory}

\subsection{Metadynamics}

Metadynamics is an innovative approach in computational science that enhances the sampling of complex molecular systems. It achieves this by dynamically adding an artificial bias to the system's energy landscape during MD simulations.
As the simulation progresses, metadynamics continuously introduces small Gaussian potentials, or ``hills'', onto the energy surface at specific collective variables $Z$ of interest. These hills counteract the system's tendency to remain trapped in local energy minima, encouraging it to explore different regions of the energy landscape. Over time, these added potentials create a ``history" that guides the system towards previously unvisited configurations.

For the collective variable $Z$, we introduce the artificial bias $B(Z)$ via

\begin{align}
    B(Z) = w \sum_i e^{-\frac{(Z-Z_i)^2}{2 \sigma^2}},
\end{align}
where $Z_i$ are the measured collective variables in the previous steps, $w$ is the potential height and $\sigma$ is the smearing parameter, determining the flatness of the bias potential. Therefore, the bias force $f_a$ for each atom $a$ is calculated by
\begin{align}
\label{eq:def:metadynamics_force}
f_a = -\frac{\partial B}{\partial Z}\frac{\partial Z}{\partial x_a}.
\end{align}

By gradually reducing energy barriers and promoting transitions between different states, metadynamics allows for the exploration of rare and critical events that might otherwise be inaccessible within the limited timescales of traditional simulations.

The method has been heavily employed for protein folding \cite{barducci2011metadynamics,granata2013characterization,mendels2018folding} in biophysics, but also in materials science for example to investigate phase transformations \cite{rogal2019neural}.

\subsection{Order parameter}

In a system of $n_{\mathrm C}$ carbon atoms, we define the order parameter $Z$ via

\begin{align}
    Z =& \sqrt{\frac{3}{2}\frac{\sum_in_i^2}{n_{\mathrm C}^2} -\frac{1}{2}},\\
    \label{eq:def:n_i}
    \text{where} \quad n_i =& \frac{\sum_{ap} \cos^2\left(\sum_kM_{pik}x_{ak}\right)}{\sum_{apj} \cos^2\left(\sum_kM_{pjk}x_{ak}\right)}.
\end{align}
$n_i$ are to be interpreted as the number  of carbon atoms occupying the sub-lattice type $i$ ($i=x,y,z$) and $x_{ak}$ is the $k$-coordinate of the carbon atom $a$ ($a=1,\dots,n_{\mathrm C}$). A site type is determined by its orientation, i.e. the direction along which the nearest iron atoms lie. We define the rotation matrices $M_{pjk}$ by

\begin{align}
    M_{0jk} = \frac{\pi}{a_0}
    \begin{pmatrix}
        0 & 1 & 1\\ 
        1 & 0 & 1\\ 
        1 & 1 & 0 
    \end{pmatrix},\quad
    M_{1jk} = \frac{\pi}{a_0}
    \begin{pmatrix}
        0 & 1 & -1\\
        -1 & 0 & 1\\ 
        1 & -1 & 0 
    \end{pmatrix}
\end{align}
where $a_0$ is the lattice parameter. This definition of $Z$ allows us to calculate the derivative with respect to the position of each atom. Therefore by using the order parameter $Z$ for the collective variable for metadynamics, we can calculate the bias forces following eq. \ref{eq:def:metadynamics_force}. This definition of the order parameter is equivalent to the one used in \cite{waseda2018ordering}, where the difference comes from the fact that in their formulation the carbon atoms were placed only in perfect octahedral sites and there was only one preferential site (i.e. $n_z > n_x = n_y$). This choice of the order parameter intrinsically captures the periodicity and the symmetry of the sublattices and therefore allows us to efficiently obtain the biased energy and forces on the fly. In the following, we consider the system to be ordered if $Z > 0.5$ and otherwise disordered.

\subsection{Mean field theory (MFT) and extension of pair-wise chemical interactions (MFT+CC)}

Following MFT, the potential energy $U$ in a system of volume $V$ is given by
\begin{align}
    \label{eq:def:U_elast}
    U_{\mathrm{elast}} = -\frac{\Delta\lambda^2(1+\nu)}{2VE}\sum_{i}n_i^2,
\end{align}
where $\Delta \lambda = \lambda' - \lambda$ is the difference of dipole moments \cite{siems1968mechanical,teodosiu2013elastic,bacon1980anisotropic,leibfried2006point} along the direction of orientation (i.e. direction of the nearest neighboring iron atoms represented by $\lambda$) and along the two other directions represented by $\lambda'$, $\nu$ is the Poisson's ratio and $E$ is the Young's modulus. In our model, we also append the short range carbon-carbon interactions by
\begin{align}
    U_{\mathrm{chem}} = \frac{1}{2}\sum_{ij}\phi_iK_{ij}\phi_j
\end{align}
where $\phi_i$ is the occupancy interstitial sites by C of the site $i$ ($i=1,\dots,N_{\mathrm{sites}}=3N_{\mathrm{Fe}}$) and $K_{ij}$ is the chemical interaction tensor, which is defined by
\begin{align}
    K_{ij} = \kappa_{ij} + c \kappa_{ij}' + c^2 \kappa_{ij}''.
    \label{eq:inter-tensor}
\end{align}
where $\kappa_{ij}$, $\kappa_{ij}'$ and $\kappa_{ij}''$ are fitting parameters. In this context, we consider $\kappa_{ij}$ as the main pairwise interaction parameters (simple pairwise) and $\kappa_{ij}'$ and $\kappa_{ij}''$ as correction terms (effective manybody). 
The entropy of the system $S$ is given by:
\begin{align}
    S = -k_{\mathrm B}\sum_i(\phi_i\ln\phi_i+(1-\phi_i)\ln(1-\phi_i))
\end{align}

The free energy $F$ is given by:
\begin{align}
    F = U - TS
\end{align}
where $U$ is only $U_{\mathrm{elast}}$ in the framework of the mean field model, and $U_{\mathrm{elast}} + U_{\mathrm{chem}}$ to account for the short range chemical interactions. Within the framework of the mean field model, the free energy can be reduced to the function of a single variable $z$ and can be solved analytically. In the classical Zener model, no temperature dependence on the thermodynamic quantities are considered. However, as shown previously \cite{waseda2018ordering}, the temperature dependence is proposed to play a significant role in the ordering behaviour. Therefore we look into the temperature dependence of the dipole tensor, atomic volume, Young's modulus and Poisson's ratio. On the other hand, the carbon concentration dependence of these thermodynamic parameters was previously found to play no significant role \cite{janssen2016influence} and therefore we omit it in this study.

\subsection{Simulation setup}
The Large-scale Atomic/Molecular Massively Parallel Simulator (LAMMPS) \cite{thompson2022lammps}, as implemented in the integrated development environment (IDE) pyiron \cite{janssen2019pyiron} was employed to conduct MD simulations, and for the measurement of thermodynamic parameters. The modified Becquart interatomic potential \cite{becquart2007atomistic,veiga2014comments} was chosen to accurately represent iron-carbon interactions. This potential has been employed previously for the investigation of Zener ordering \cite{sinclair2010ordering,waseda2018ordering} and is known to reproduce qualitatively the ordering behaviour measured experimentally \cite{xiao1995lattice}. The simulation box contained 2000 iron atoms, with carbon concentrations ranging from 1\% to 11\%. Temperatures spanned 100~K to 1000~K, covering a wide thermal range, where the box shape was allowed to vary to keep the box free of pressure.

To explore the impact of initial configurations, 20 simulation boxes were utilized per temperature. Ten of these boxes began with ordered atomic arrangements, while the other ten had disordered configurations, where it was made sure that no two atoms are within a distance of one lattice parameter in order to avoid an unrealistic starting condition. Simulations were conducted until the convergence of the free energy histogram was reached, but at least for 1~ns to capture thermal effects accurately.

For metadynamics histogram parameters, we employed the smearing value of $\sigma=0.01$. The histogram $B(Z)$ was updated every 100 steps with the value of $w=0.1$~meV. The histogram was stored in the form of a 1-D mesh grid with 1,000 points, meaning within one $\sigma$, there were 10 grid points. At each integration step, the metadynamics forces on the carbon atoms are calculated according to eq.~\ref{eq:def:metadynamics_force}. In order to prevent the collective motion of the system, the counteracting forces are added to the iron atoms. In particular, the sum of all metadynamics forces divided by the number of iron atoms is added to the force of each of the iron atoms.

In parallel, we also ran conventional MD simulations for the temperature range of 600~K to 1,400~K for 1~ns and calculated the order parameter based on the final configuration.

The temperature dependence of the thermodynamic parameters was measured by running MD calculations for various temperatures for 100~ps. The necessary values (pressures, energy etc.) were measured by taking the average of all steps after the initial 10~ps.

\section{Results}

\subsection{Temperature effects on thermodynamic parameters}
In order to include the vibrational entropy into MFT, we first studied the temperature effects on the following thermodynamic parameters: difference of dipole moments $\Delta \lambda$, Young's modulus, thermal expansion and Poisson's ratio.
\begin{figure}
 \centering
  \includegraphics[width=\columnwidth]{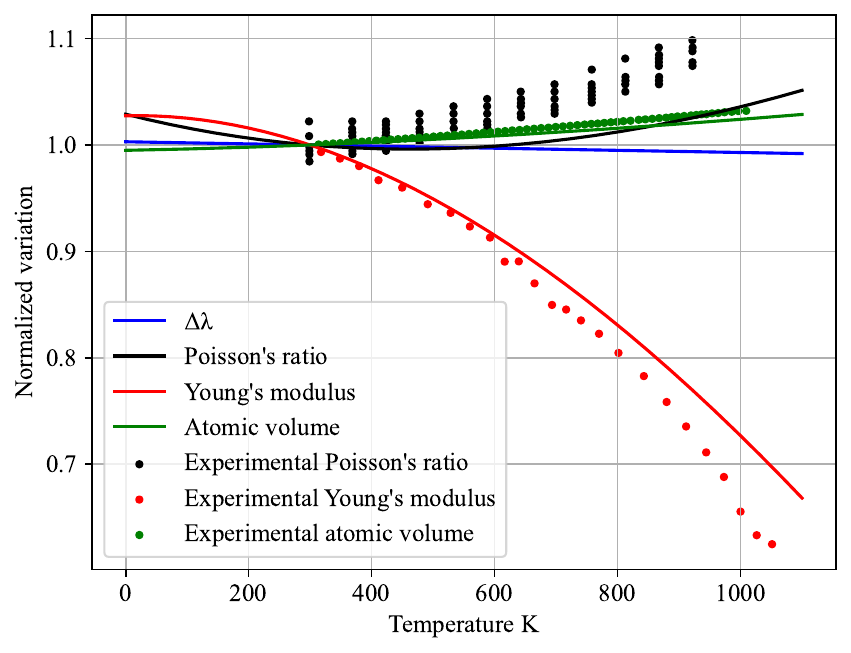}
 \caption{Temperature dependence of thermodynamic parameters normalized to the value at 300~K. Experimental results were taken from \cite{batista2014development} for the Poisson's ratio, \cite{fukuhara1993elastic} for the Young's modulus and \cite{denand2020carbon} for the atomic volume.}
 \label{fig:thermo_param_temp_dep}
\end{figure}

Fig.~\ref{fig:thermo_param_temp_dep} displays the variation of thermodynamic parameters as a function of the temperature. The comparison with the experimental results shows a good agreement. It can be clearly seen that the dipole moment hardly changes with the temperature. There is some slight change for the Poisson's ratio, but as the change enters with the factor of $1 + \nu$ in eq.~\ref{eq:def:U_elast}, this temperature change as well as the deviations from experiment can be safely ignored, as it lies around a value of 0.3. The temperature dependence of the Young's modulus, on the other hand, has an important impact on the hardness of the material, and we can expect the elastic energy to have a higher weight as the temperature increases compared to the case where the 0~K Young's modulus is used for the entire temperature spectrum. This effect is somewhat weakened by the increase in volume with the increasing temperature.

\begin{figure}
 \centering
  \includegraphics[width=\columnwidth]{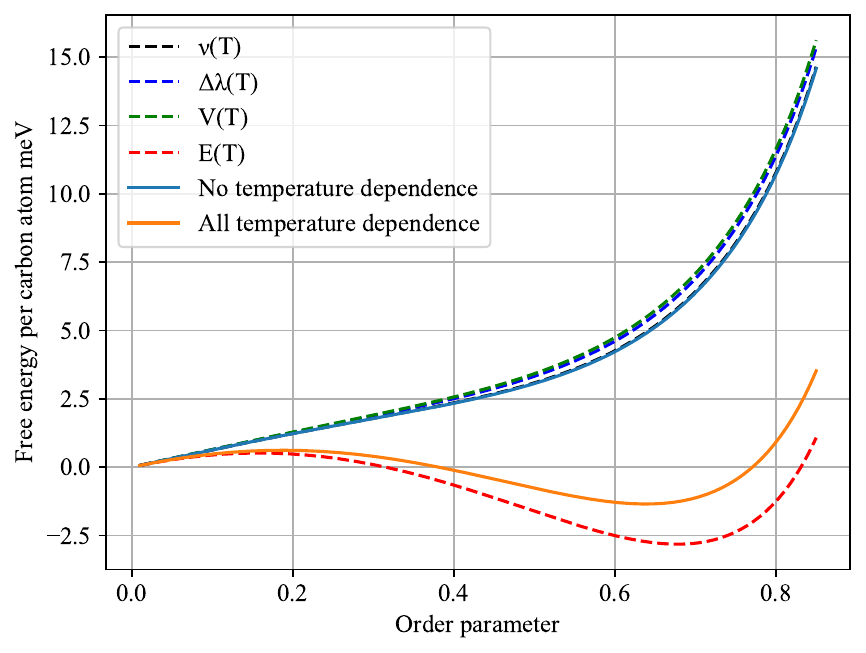}
 \caption{Free energy as a function of order parameter $Z$ for 7.5~at.\%C at 900~K with temperature dependence for Poisson's ratio $\nu$, difference of dipole moments $\Delta\lambda$, Young's modulus $E$ and volume $V$ individually, as well as no temperature dependence (i.e., values measured for 0 K) and all of them together.}
 \label{fig:mean_field_free_energy}
\end{figure}

Fig.~\ref{fig:mean_field_free_energy} shows the free energy variation as a function of the order parameter for 7.5~at.\%C at 900~K. For the four parameters that have been investigated, we can see that only the Young's modulus has a considerable effect on the free energy. This figure also shows that the softening of iron becomes so important that the system becomes ordered by considering it, as the free energy minimum falls in the region $Z > 0.5$ in this case.

\subsection{Short range chemical interactions}

Next, in order to understand the short-range carbon-carbon interactions, we inserted two carbon atoms in a box with 2,000 iron atoms for all possible configurations, for which a structure optimization was performed to obtain the minimum energy. The interaction between two carbon atoms $i$ and $j$ is calculated via

\begin{align}
    \kappa_{ij} = E(C_i, C_j) - 2 E(C) + E(\varnothing) + \Delta E_{ij},
\end{align}
where $E(C_i, C_j)$ is the total energy of the box containing two carbon atoms at positions $i$ and $j$, $E(C)$ is the total energy of the box with one carbon atom and $E(\varnothing)$ is the box energy without carbon atoms. The additional value $\Delta E_{ij}$ is calculated by (cf. eq.~\ref{eq:def:U_elast})

\begin{align}
    \Delta E_{ij} =
    \begin{cases}
        \frac{\Delta \lambda^2 (1+\nu)}{VE},& \text{if same orientation}\\
        0, & \text{otherwise}
    \end{cases}
\end{align}
$\Delta E_{ij}$ is needed to counteract double counting of long-range elastic interactions. Two carbon atoms share the same orientation, when their respective nearest iron atoms lie along the same direction.

\begin{figure}
 \centering
  \includegraphics[width=\columnwidth]{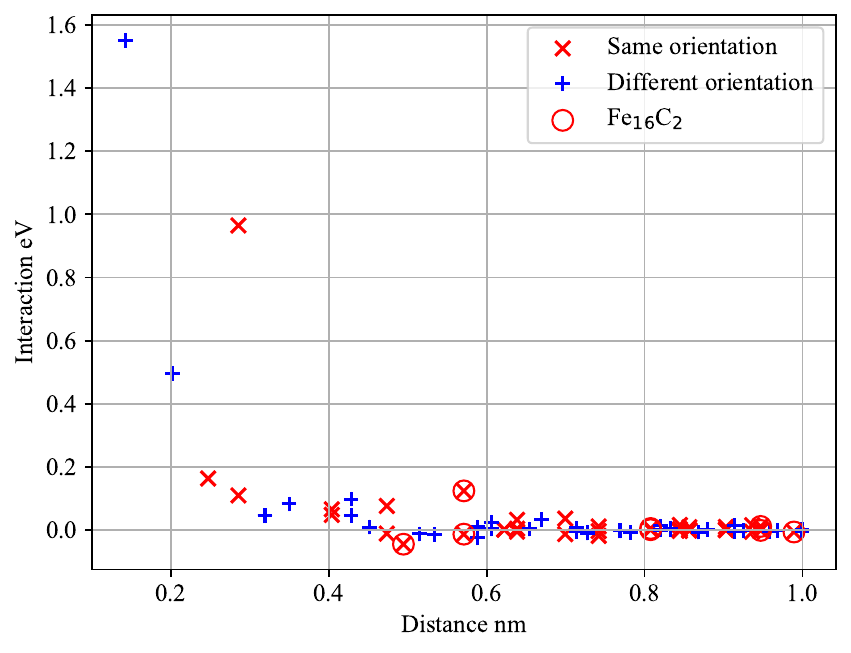}
 \caption{Chemical interactions between two carbon atoms as a function of the distance. Distances corresponding to the Fe$_{16}$C$_2$ configuration are marked with circles additionally. Positive values are repulsive interactions.}
 \label{fig:CC_chemical_interactions}
\end{figure}

As previously also shown \cite{ruban2014self,domain2004ab}, the direct carbon-carbon interactions are mostly repulsive, as shown in fig.~\ref{fig:CC_chemical_interactions}. There are a few exceptional points, most notable for the ones which would correspond to the Fe$_{16}$C$_2$ configuration. As a matter of fact, the interaction between two carbon atoms is most attractive for this potential, when two carbon atoms sit at a distance of one lattice parameter in each direction, which is the shortest distance between two carbon atoms in the Fe$_{16}$C$_2$ configuration.

In order to compute $\kappa_{ij}'$ and $\kappa_{ij}''$ in eq.~\ref{eq:inter-tensor}, we distributed carbon atoms via simple Monte Carlo using $\kappa_{ij}$ for various temperatures and carbon contents, measured the energies and fitted them using the Ridge regression by minimizing the following cost function:

\begin{align}
    \mathcal{L} = \sum_k (U(\phi_k) - \sum_{ij}\phi_{k,i} K_{ij} \phi_{k,j})^2 + \lambda\sum_{ij}(\kappa_{ij}'^2 + \kappa_{ij}''^2),
\end{align}
where
\begin{align}
    U(\phi) = E(\phi) - n_{\mathrm C}E(C) + (n_{\mathrm C} -1) E(\varnothing)
\end{align}
and $\phi_k$ is the occupancy vector (i.e. filled with 0 and 1, where 0 is empty and 1 is filled), $\lambda$ is the regularization parameter of the Ridge regression, and $E(\phi)$ is the box energy of the occupancy $\phi$. The cost function $\mathcal L$ is minimized with respect to $\kappa_{ij}'$ and $\kappa_{ij}''$. In order to see the quality of fitting, we created a data set of 2,000 points, where we set 25~\% of the points aside for the test set. We estimated the value of $\lambda$ using the Leave-One-Out (LOO) using the remaining 75~\% of the data points.

\begin{figure}
    \centering
    \includegraphics[width=\columnwidth]{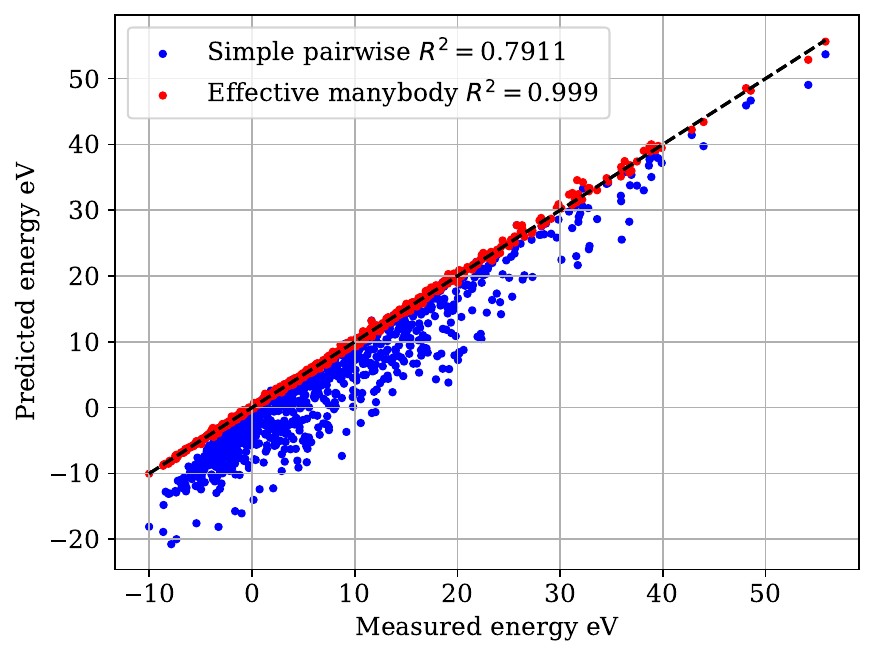}
    \caption{Energy prediction using the pairwise interactions without correction (simple pairwise, without $\kappa_{ij}'$ and $\kappa_{ij}''$) and with correction (effective manybody, with $\kappa_{ij}'$ and $\kappa_{ij}''$) for the test set}
    \label{fig:comp_fitting_with_corr}
\end{figure}

As it turned out, the superposition of the pairwise interactions with only $\kappa_{ij}$ for boxes containing more than 2 carbon atoms the energy was generally underestimated (\emph{cf.} fig.~\ref{fig:comp_fitting_with_corr}). Despite the lack of explicit many-body interactions, the energies with more than 2 atoms can be correctly reproduced with the correction terms.

\begin{figure}
    \centering
    \begin{tabular}{c}
        MFT vs. MFT+CC vs. MMC\\
        \includegraphics[width=\columnwidth]{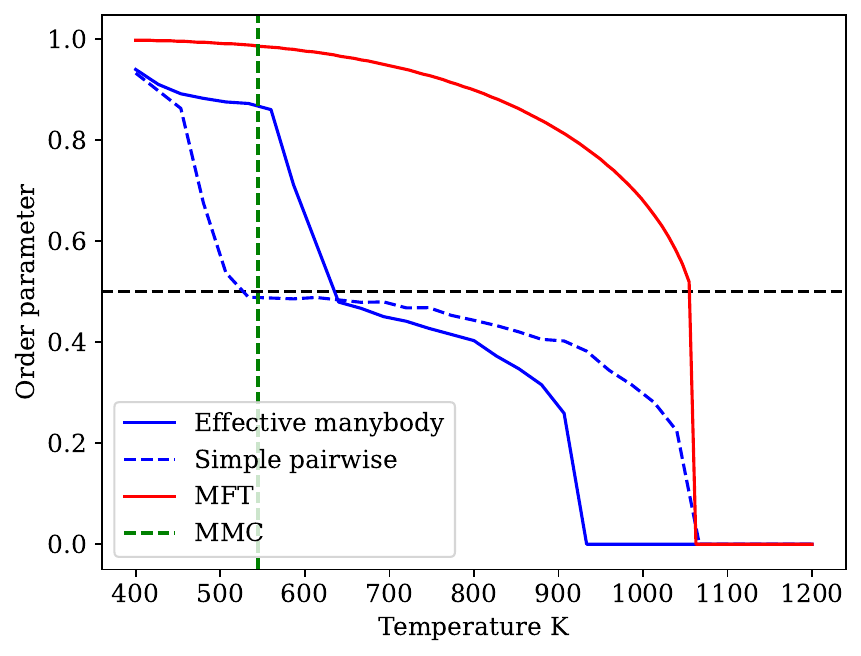}
    \end{tabular}
    \caption{Comparison of MMC results and MFT+CC results (one with simple pairwise interactions, the other with the effective many-body interactions) at 8~at.\%C. For a comparison, the MFT and MMC results \cite{waseda2018ordering} are also shown.}
    \label{fig:pd_comp_MC_MFT_8pct}
\end{figure}

Using these chemical interactions, we obtained the order-disorder transition behaviour at 8~\%at.C as shown in fig.~\ref{fig:pd_comp_MC_MFT_8pct} for MFT+CC. The results are compared against the classical MFT and the results from MMC \cite{waseda2018ordering}, which intrinsically has long and short range carbon-carbon interactions, but does not have the temperature effects beyond configurational entropy. As it was speculated (cf. \cite{waseda2018ordering}), the addition of direct carbon-carbon interactions decreases the transition temperature at high carbon concentration in MFT+CC compared to MFT, and it shows a significantly better agreement with the MMC results, which also contain direct carbon-carbon interactions. The simple pairwise interactions gave ambiguous behaviour, staying somewhere between ordering and disordering at lower temperatures, and only slowly converging to the disordered state around the same temperature as MFT. We have observed that there are configurations for two carbon atoms which deliver particularly low energies. These values got corrected with effective many-body interactions.
\begin{figure}
    \centering
    \begin{tabular}{c}
        MFT vs. MFT+CC vs. MMC @ 1~at.\%C\\
        \includegraphics[width=\columnwidth]{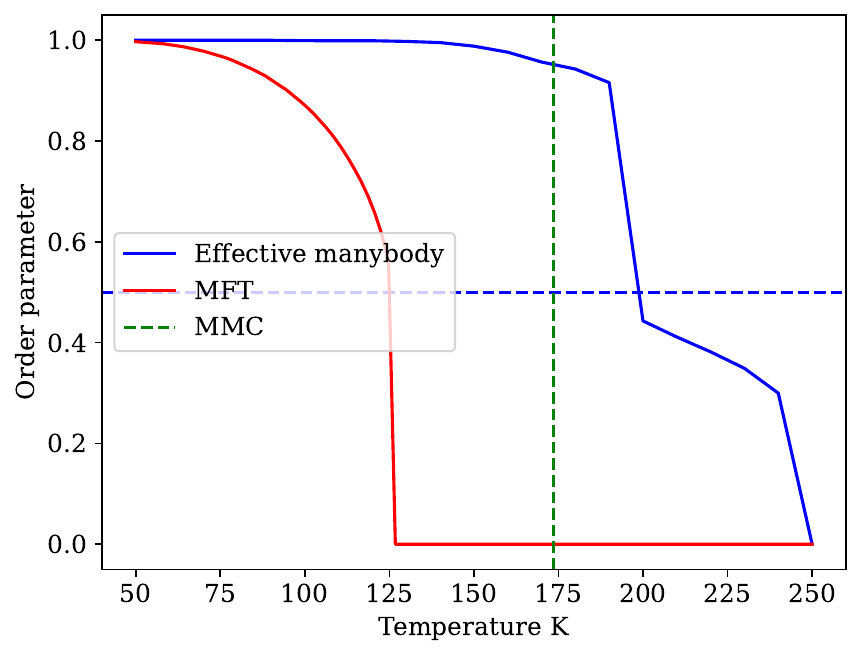}
    \end{tabular}
    \caption{Comparison of MMC results \cite{waseda2018ordering}, MFT+CC and MFT at 1~at.\%C}
    \label{fig:pd_comp_MC_MFT_1pct}
\end{figure}

Previous results \cite{waseda2018ordering} had indicated the presence of the Fe$_{16}$C$_2$ structure, which was predicted to be the reason why MMC showed a higher transition temperature than MFT. In our study, we could observe the same behaviour at 1~\%at.C (fig.~\ref{fig:pd_comp_MC_MFT_1pct}). In order to better understand the sublattice distribution, we plotted the site radial distribution in fig.~\ref{fig:rdf_1pct_175K}, which we defined by:

\begin{figure}
    \centering
    \includegraphics[width=\columnwidth]{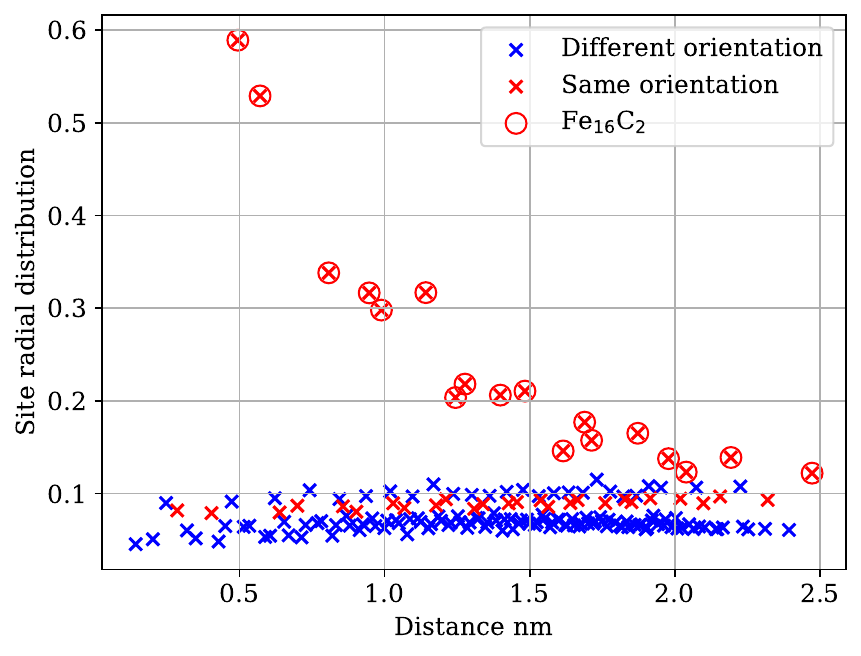}
    \caption{Site radial distribution as defined in eq.~\ref{eq:SRD} for MFT+CC and for 1~\%at.C at 175~K}
    \label{fig:rdf_1pct_175K}
\end{figure}

\begin{align}
    \rho(r_k) = \sum_{i>j}\frac{\sqrt{\phi(r_i)\phi(r_j)}}{n_k|\phi|}\delta(r_k-|r_i-r_j|)
    \label{eq:SRD}
\end{align}
where $n_k$ is the number of site pairs for the shell $k$ and $|\phi|$ is the average occupation. We could indeed observe that at the transition temperature predicted by MMC, there is a strong preference for the sites corresponding to Fe$_{16}$C$_2$. As a matter of fact, we can also observe that for all other pairs, there is no strong preference, regardless of whether the site types are the same or not, indicating without Fe$_{16}$C$_2$, the system could very well be disordered.

\begin{figure}[htb]
    \centering
    \includegraphics[width=\columnwidth]{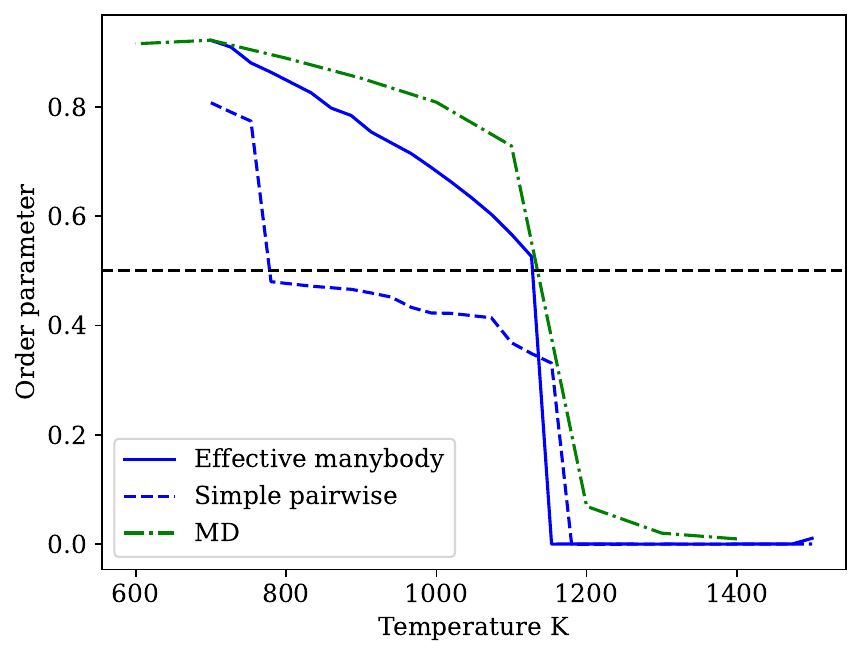}
    \caption{Comparison of MD results and MFT+CC+FT at 8~\%at.C for the order parameter as a function of the temperature}
    \label{fig:pd_comp_MD_MFT}
\end{figure}

We now add the temperature dependence of thermodynamic parameters in MFT+CC to make MFT+CC+FT and compare the results against classical MD results in fig~\ref{fig:pd_comp_MD_MFT}, which intrinsically have both long and short range carbon-carbon interactions and all temperature effects. Firstly, we can observe the significant increase of the transition temperature in both MFT+CC+FT and MD, compared to the previous results for MFT+CC and MMC in fig.~\ref{fig:pd_comp_MC_MFT_8pct}. Quite crucially, the simple pairwise interactions deliver a much lower order parameter in the area, in which the carbon atoms are ordered according to the MD results. This behaviour is corrected in the effective many-body interactions, following the MD results almost perfectly.

\subsection{Metadynamics and comparison of all methods}
Finally, we compare the foregoing results with the ones obtained from metadynamics, which are comparable to MD results, but can explore the free energy surface along the order parameter for any temperature and any carbon content.
\begin{figure}
    \centering
    \begin{tabular}{c}
        MFT vs. Metadynamics @ 8~at.\%C\\
        \includegraphics[width=\linewidth]{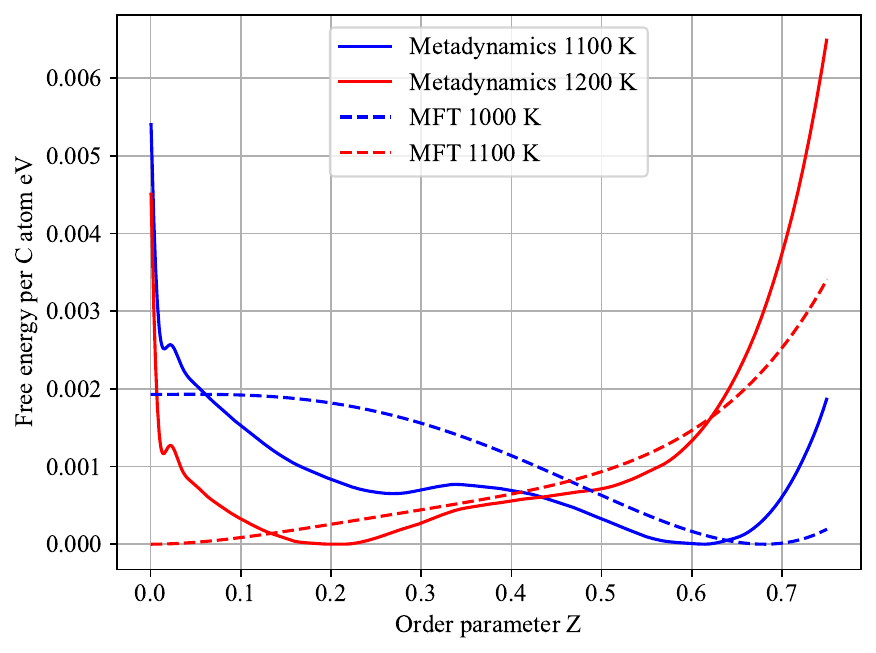}
    \end{tabular}
    \caption{Free energy measured with MFT and metadynamics with 8~at.\%C, relative to the respective minimum.}
    \label{fig:MF_metadynamics_free_energy_comparison}
\end{figure}

Fig.~\ref{fig:MF_metadynamics_free_energy_comparison} shows the free energy surface for 8.0~at.\%C across the transition temperature for metadynamics, and also for MFT for a comparison. In both cases we can see a clear first order behaviour with qualitatively similar values, except for the lower end of the minimum, which is $z = 0$ for MFT and somewhere around $z = 0.25$ for metadynamics at higher temperature (1100~K for MFT and 1200~K for Metadynamics). It could be related to the fact that in MFT there are only discrete octahedral sites to take, while in metadynamics any position of any C atoms has its contribution (cf. eq.~\ref{eq:def:n_i}), so that thermal vibrations of C atoms lead to a slight deviation from $z = 0$. Similarly, Metadynamics has a strong increase at higher order parameter compared to MFT.

\begin{figure}
    \centering
    \includegraphics[width=\columnwidth]{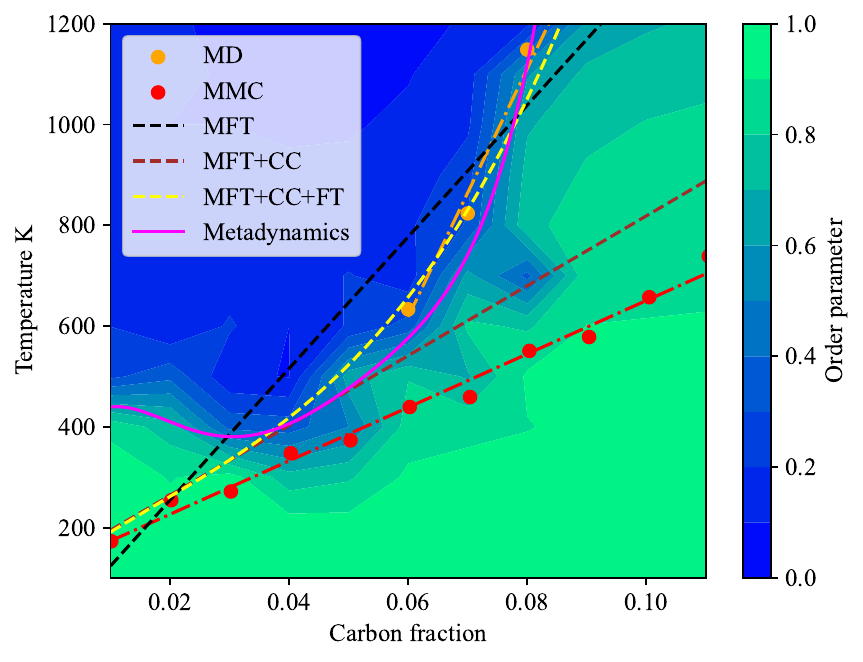}
    \caption{Phase diagram of order-disorder transition for Metadynamics (background and phase boundary interpolated by a magenta line) in comparison with other methods discussed in previous sections}
    \label{fig:pd_metadynamics_final}
\end{figure}

Finally, we show the phase diagram delivered by metadynamics in fig.~\ref{fig:pd_metadynamics_final} with the transition temperature curves from previous sections. We can divide this diagram in three parts: low carbon range ($<$2~at.\%C), intermediate range ($>$2~at.\%C \& $<$7~at.\%C) and high carbon range ($>$7~at.\%C). In the low carbon range, models including carbon-carbon interactions (MFT+CC, MFT+CC+FT, MMC and metadynamics) show ordering behaviour, which cannot be explained within the framework of MFT. Then in the intermediate range, there is a suppression of ordering, which is observed in all models containing short range carbon-carbon interactions, i.e. all but MFT. Then at high carbon range, there is a strong increase of the transition temperature, which is observed by models containing vibrational entropy (MFT+CC+FT, MD and metadynamics), which is suppressed in MMC and MFT+CC. MFT, on the other hand, goes through the middle of the phase diagram, by coincidence reaching more or less the same area as the results from metadynamics, as the combination of all effects roughly nullifies the overall effect. We found that with metadynamics the transition temperature at very low C range significantly higher than with MMC or MFT+CC+FT. It may contain some effects that we did not consider in this article, but one reason for this is because at lower temperatures the fluctuations in the energy landscape were a lot larger than the energy difference between ordered and disordered states, so that it was extremely difficult to have fully converged results. Therefore we may see a different result if we increase the number of calculations significantly.

\section{Conclusion and discussion}

In this article, we did a comparison of order-disorder transition behaviour using different methods, each of which captures certain thermodynamic properties others do not offer. In the initial phase of this study, our focus was directed towards a comprehensive examination of the thermodynamic parameters that constitute the foundational framework of the mean-field theory (MFT). Notably, this investigation revealed a pronounced modulation in the Young's modulus as a function of temperature. This observation implies that the material exhibits a propensity for softening at elevated temperatures. In contrast, all other thermodynamic parameters entering MFT showed small/negligible changes in temperature. Including short-range carbon-carbon interactions in MFT in form of pairwise interactions (MFT+CC)  lowered the order-disorder transition temperature. This tendency is in agreement with the previous MMC results, which also contain short-range carbon-carbon interactions. By including the explicit temperature dependence of the thermodynamic parameters (MFT+CC+FT), we could show a good agreement with the MD results in transition temperature and explain the difference between MMC and MD quantitatively. Finally, we employed metadynamics to explore all areas of the phase diagram and showed a good agreement with MFT+CC+FT.

The findings of this work have the following important implications: Firstly, it shows that the order-disorder transition is difficult to observe in supersaturated iron until very high temperatures are reached, since the direct carbon-carbon interactions suppress the ordering. This might indeed be the reason why there are not many experimental results that attest to the actual order-disorder transition. Secondly, while we did not handle the issue in this article, we could observe a strong dependence of the transition behaviour on the short-range carbon-carbon interactions by varying their values slightly. This is because long-range elastic interactions are a lot weaker than the short-range interactions, and therefore it is possible that by employing a different model (e.g. DFT etc.) it forms a carbide rather than converging to a Zener ordering, which was indeed also been witnessed in previous studies \cite{huang2023atomistic}. At the same time, we could also see that both linear elasticity and pairwise interactions are limited to delivering qualitative understanding when it comes to intricate interactions of carbon atoms, as they are both outstandingly non-linear and of many body nature, in addition to the finite temperature effects shown in this article. Lastly, the combination of these factors, namely the sensitivity to the temperature dependence and many-body nature of the carbon interactions is inherently very difficult to address to using conventional atomistic methods such as DFT, especially if more factors such as other chemical elements or presence of defects have to be taken into account.

\section*{Supplementary material}
All materials used to produce this article can be found on this page: \url{https://github.com/samwaseda/zener_ordering}

\section*{Acknowledgement}
We gratefully acknowledge the financial support from the German Research Foundation (DFG) under grant HI 1300/15-1 within the DFG-ANR project C-TRAM.

\bibliography{cite}

\end{document}